\documentclass{elsart}

\usepackage{natbib}

\usepackage{graphicx}

\usepackage{amssymb}
\newcommand{\micro}{\mbox{\usefont{U}{eur}{m}{n}\char22}}

\begin{document}

\begin{frontmatter}



\title{Blazar observations above 60~GeV: the Influence of CELESTE's Energy Scale on the Study of Flares and Spectra}


\author{Brion, E.}
\address{CENBG, Domaine du Haut-Vigneau, BP 120, 33175 Gradignan Cedex, France}
\ead{brion@cenbg.in2p3.fr}
\author{and the CELESTE collaboration}

\begin{abstract}
The CELESTE atmospheric Cherenkov detector ran until June 2004. It has observed the blazars Mrk~421, 1ES~1426+428 and Mrk~501. We significantly improved our understanding of the atmosphere using a LIDAR, and of the optical throughput of the detector using stellar photometry. The new data analysis provides better background rejection. We present our light curve for Mrk~421 for the 2002-2004 season and a comparison with X-ray data and the 2004 observation of 1ES~1426+428. The new analysis will allow a more sensitive search for a signal from Mrk~501.
\end{abstract}

\begin{keyword}
CELESTE\sep Cherenkov\sep Mrk~421\sep 1ES~1426+428\sep Mrk~501
\PACS 95.85.Pw 
\sep 98.54.Cm  
\end{keyword}

\end{frontmatter}

\section{Introduction}

CELESTE was a Cherenkov experiment using 53 heliostats of the former \'Electricit\'e de France solar plant in the French Pyrenees at the Th\'emis site. It detected Cherenkov light from electromagnetic showers produced in the atmosphere by the $\gamma$-rays coming from high energy astrophysical sources. The light is reflected to secondary optics and photomultipliers installed at the top of the tower. Finally it is sampled to be analysed (Par\'e 2002).

To constrain the energy scale of the experiment we improved the optics simulation, now in good agreement with the data. The data analysis has also been improved so that we have better background rejection. We present the light curve for Mrk~421.

\section{Constraining the energy scale} \label{sec:simulation}

\begin{figure}[htbp]
   \centering\includegraphics[width=7.0cm]{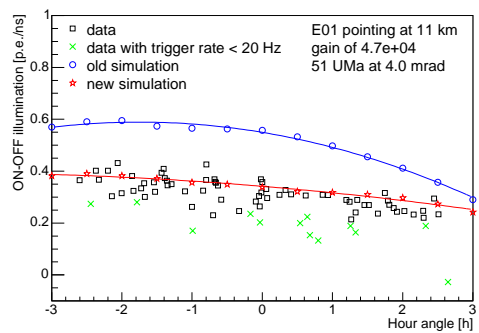}
   \caption{ON$-$OFF illumination from the star 51~UMa (M$_\mathrm{B}=6.16$) in the FOV of Mrk~421 as a function of hour angle for pointing at 11~km: the new simulation with our corrections (red stars) fits the data well (black squares) whereas the old simulation (blue circles) was 50~\% too high.}
   \label{fig:simulation}
\end{figure}

The simulation has been reexamined to reduce the uncertainties on the energy scale of the experiment (Brion 2003). The LIDAR operating on the site for atmospheric monitoring provided a better determination of the atmospheric extinction (Buss\'ons Gordo 2004). A stellar photometry study, focussing on the comparison between simulations and data on bright stars' currents, has been done on the star 51~UMa (M$_\mathrm{B}=6.16$) which is in the field of view (FOV) of Mrk~421. This showed that the old simulation was too optimistic. All mirror reflectivities were decreased subsequent to new measurements. The nominal focussing of the heliostats was degraded after a study of star image sizes. We also verified the photomultiplier gains. The results for the old data set (40~heliostats, new is 53~helisotats, see~\S~\ref{sec:analysis}) are presented in figure~\ref{fig:simulation}: the effect of these changes is smaller for $\gamma$-ray showers (an extended light source) than for stars (point sources).

This study helped us to perform our selection criteria for the data: all data that are too low in currents on the star have also low trigger rates. In order to trigger CELESTE, the heliostats, except the \textit{veto} (see \S~\ref{sec:veto}), are split into 6~groups. For each of them, the analog sum of signals gives the first level trigger. Then, a logical pattern is defined on the majority of the triggering groups. Thus, in the case of the source Mrk~421, we reject all data with trigger rates under 20~Hz for the old data set and under 16~Hz for the new data set.

\begin{figure}[htbp]
\begin{center}
   \includegraphics[width=6.0cm]{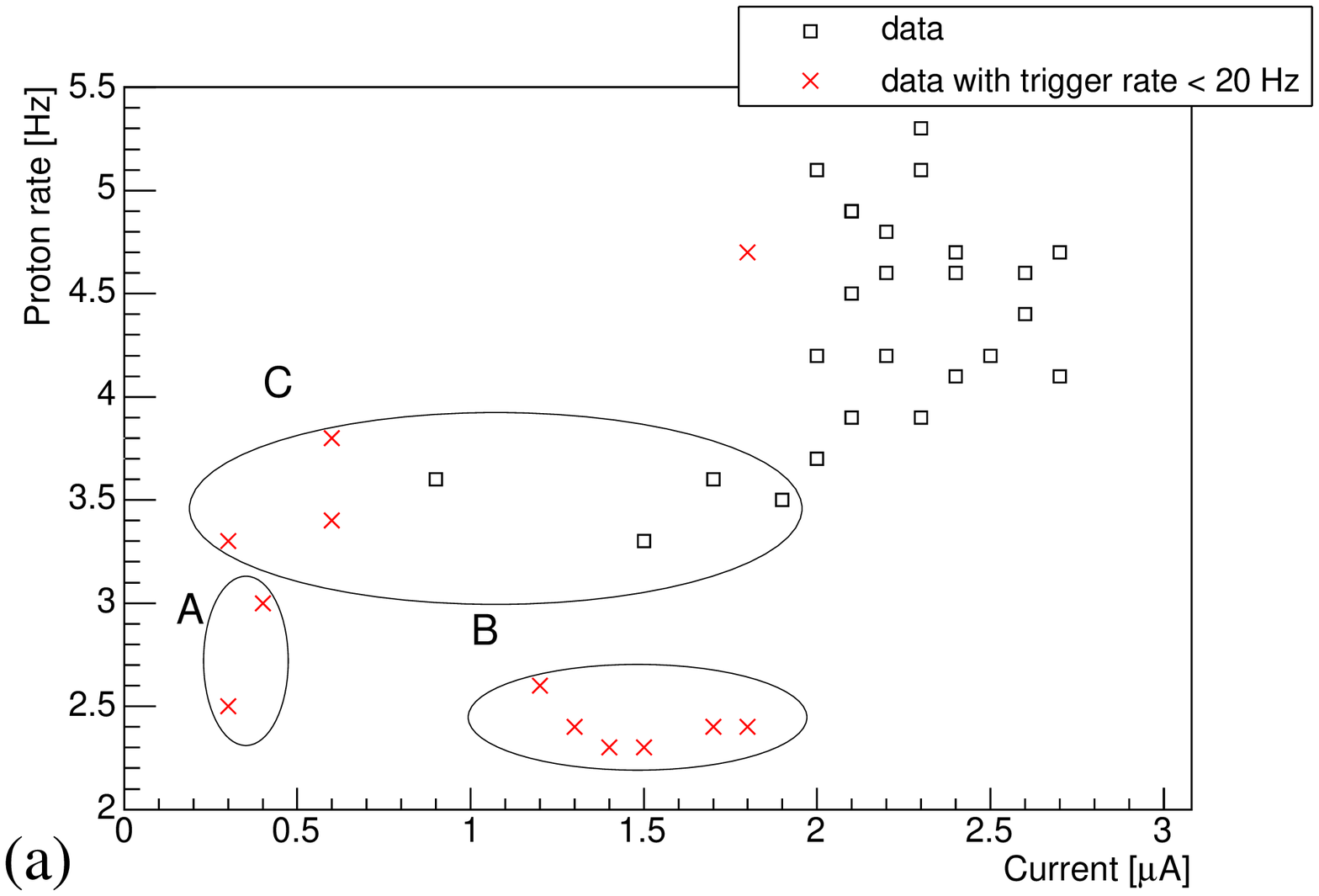}
   \includegraphics[width=6.0cm]{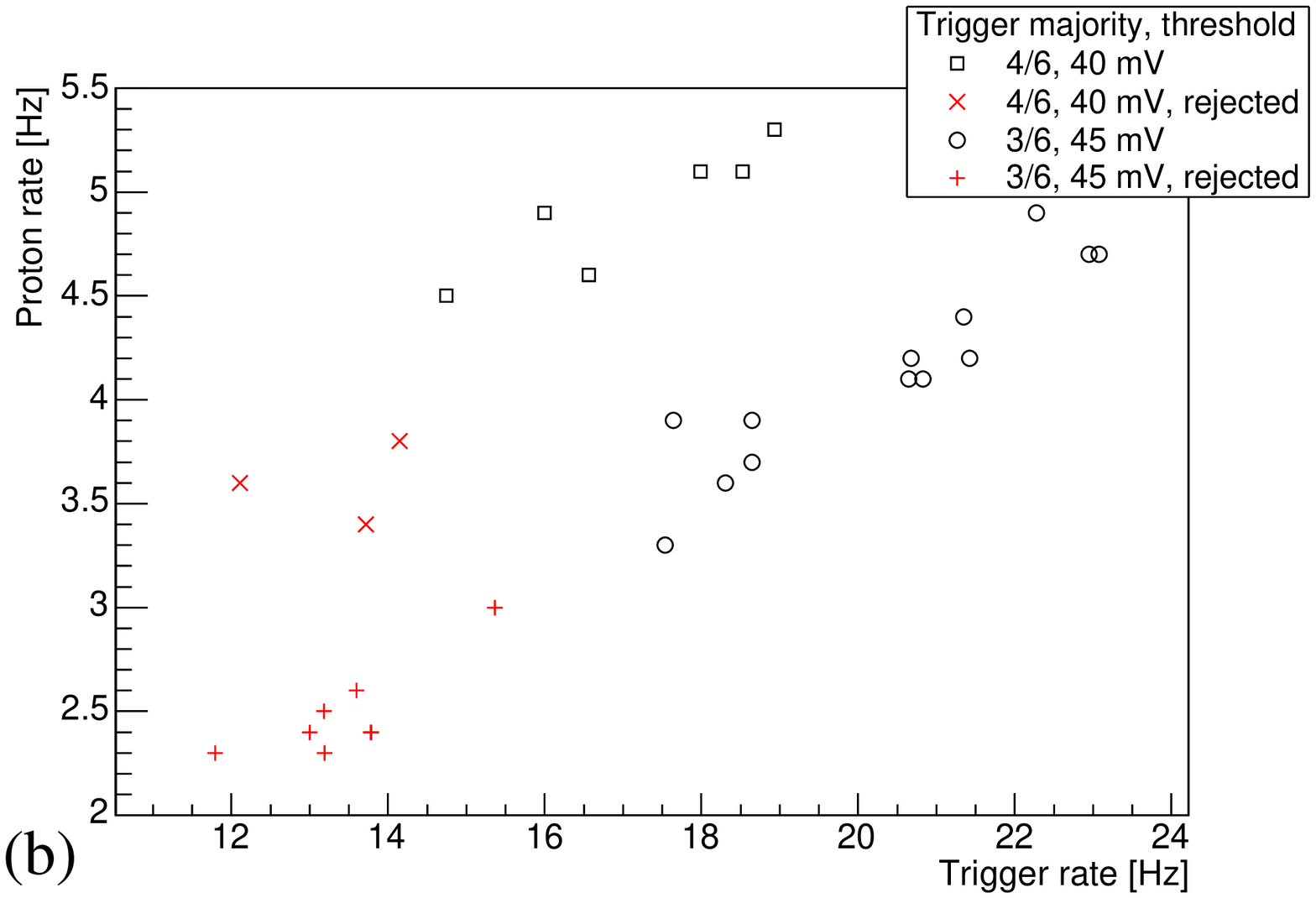}
\end{center}
   \caption{(a) Proton rate as a function of ON$-$OFF current for heliostat E03 that sees 51~UMa when pointing Mrk~421. For a typical photomultiplier tube gain of $5.6\times10^{4}$, $2.5$~\micro A corresponds to 0.28~p.e./ns, as seen in figure~\ref{fig:simulation}. (b) Proton rate as a function of trigger rate for data on Mrk~421.}
   \label{fig:ProtonRate}
\end{figure}

We also looked at the proton rates as a standard candle for the detector which should be stable for good quality nights. These rates are determined with high offline threshold cuts to avoid trigger bias, and are therefore low (typical trigger rates $\sim 22$~Hz). We've shown that they are correlated with the currents on star 51~UMa for selected data (new data set, figure~\ref{fig:ProtonRate}~(a)). The data with low rates are now still rejected but perhaps some of them could be corrected. Indeed, three doubtful zones, A, B and C on the figure, can be distinguished. The zone A can be interpreted as bad nights with thick cloud cover (weak star, little Cherenkov light), the zone B as nights with aerosols and above average extinction of Cherenkov and starlight, and the zone C as nights with high clouds that stop starlight but not Cherenkov light. The data in this last zone may therefore be used. Figure~\ref{fig:ProtonRate}~(b) shows the correlation between the proton rate and the trigger rate for the same data set. Defining a selection criteria for each type of acquisition, and based only on these rates, would be very interesting for sources that don't have any star nearby for a photometry study.

\section{Analysis improvement} \label{sec:analysis}

Since the 2000 status of the experiment (de Naurois 2002) three main changes were made for the analysis. First, the selection criteria of the data are stricter for the current and trigger rate stability, and as we've shown for the trigger and proton rate value (work in progress). The experiment has been upgraded from 40 to 53 heliostats. We use part of them to broaden our narrow FOV. Finally, we have found a new method to exploit the FADC information to reject the hadronic background (Manseri 2004 a, b). The second and third point are developed hereafter.

\subsection{The \textit{veto} configuration} \label{sec:veto}

\begin{figure}[htbp]
  \begin{minipage}[t]{0.45\textwidth}
    \centering\includegraphics[width=\linewidth]{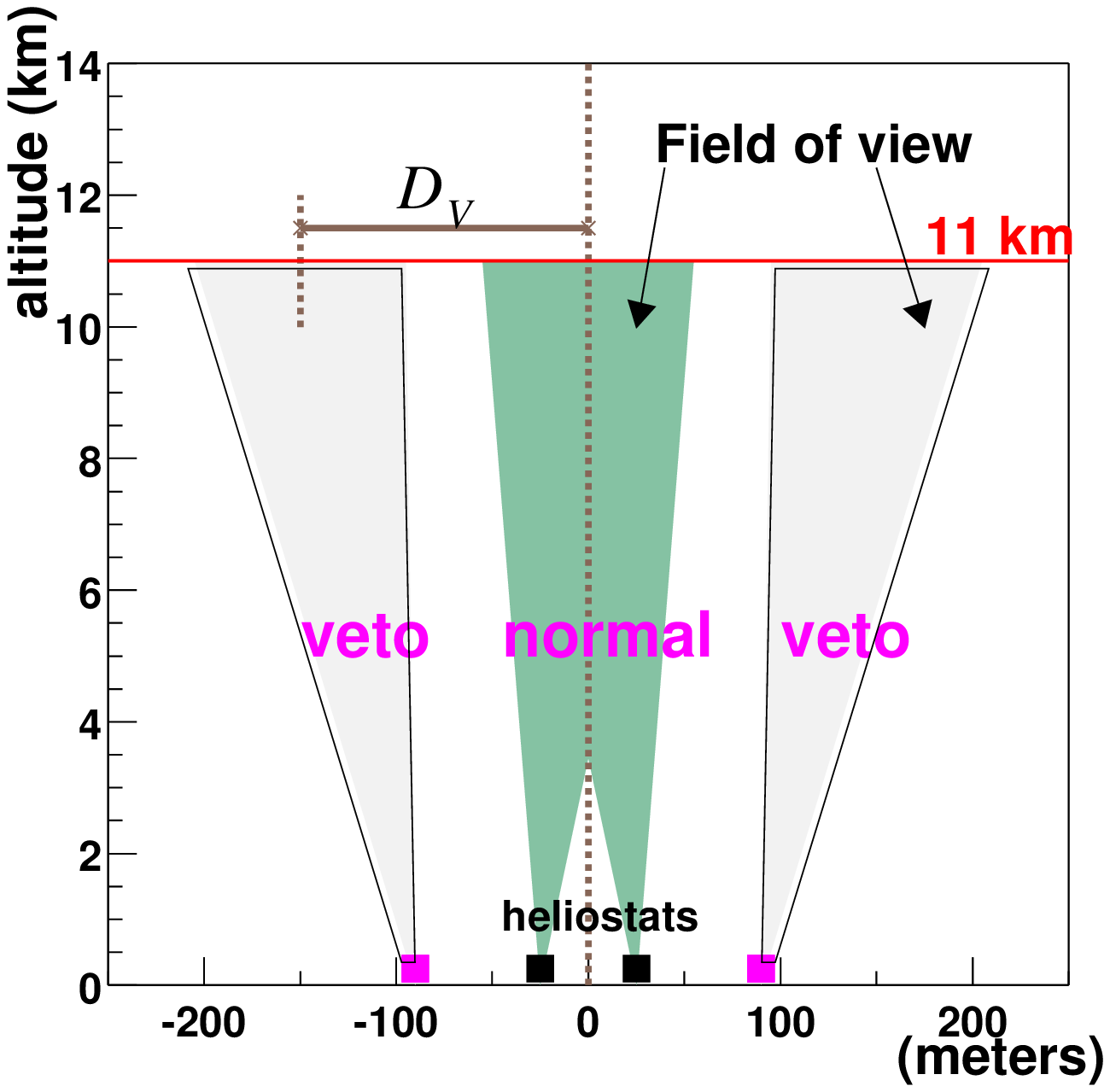}
    \caption{Scheme of the \textit{veto} configuration.}
    \label{fig:PrincipeVeto}
  \end{minipage}
  \hfill
  \begin{minipage}[t]{0.45\textwidth}
    \centering\includegraphics[width=\linewidth]{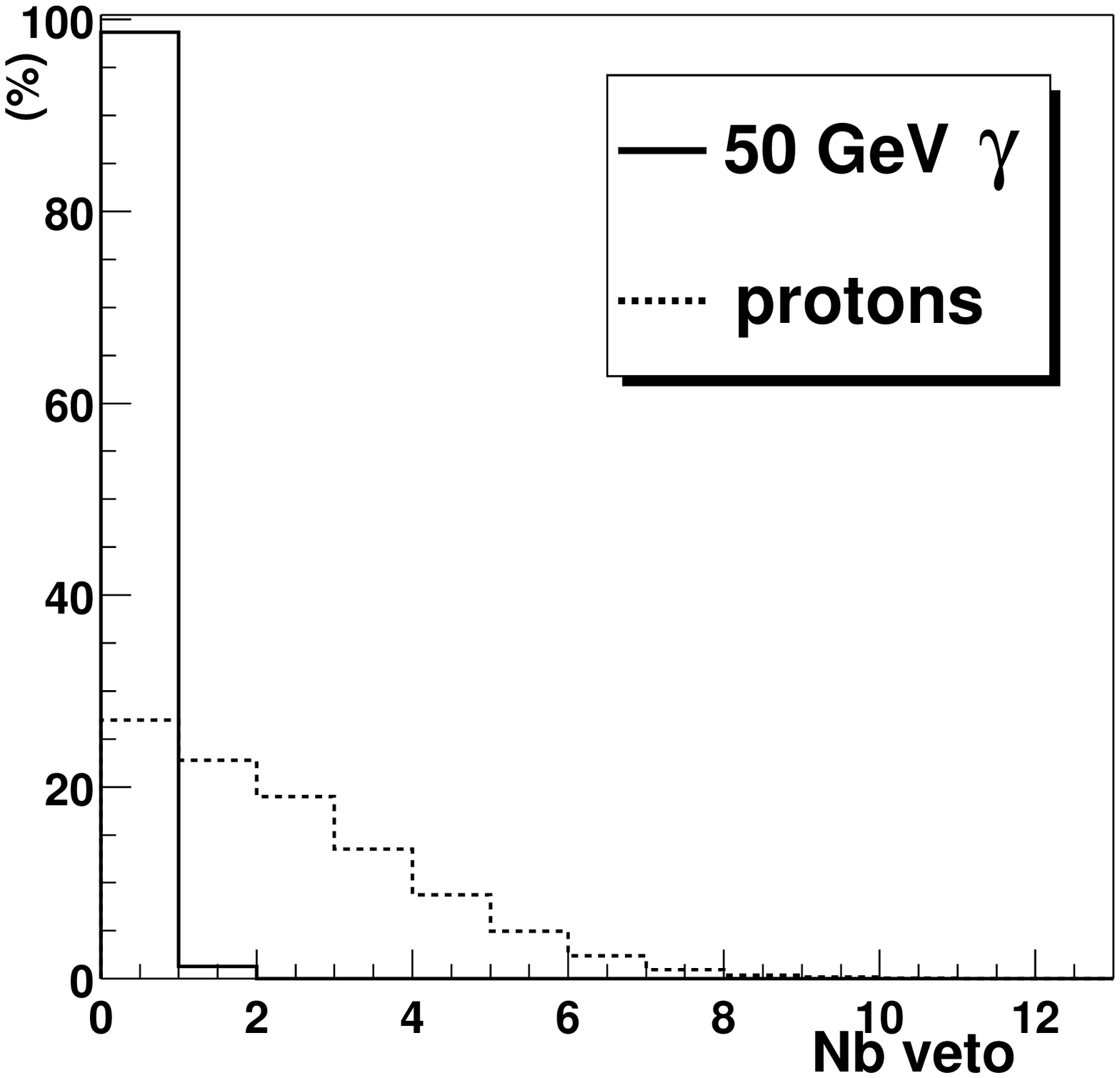}
    \caption{Distribution of the number of illuminated \textit{veto} heliostats (simulations).}
    \label{fig:NbVetoMC}
  \end{minipage}
\end{figure}

Hadronic showers have a more chaotic and extended development than electromagnetic showers. To measure the extent of the shower, we artificially broaden the FOV: as before, all heliostats aim at 11~km above the ground in the direction of the source, where the maximum of the shower is supposed to occur in our energy range. But 12~heliostats, distributed around the edge of the field, sample a ring of 150~m around that point (figure~\ref{fig:PrincipeVeto}).

Because of the compactness of electromagnetic showers, the light does not illuminate these 12~heliostats, named {\it veto}, contrary to hadronic showers (fi\-gure~\ref{fig:NbVetoMC}). So we require that no \textit{veto} be illuminated. 

\subsection{FADC information}

\begin{figure}[htbp]
  \centering\includegraphics[width=\textwidth]{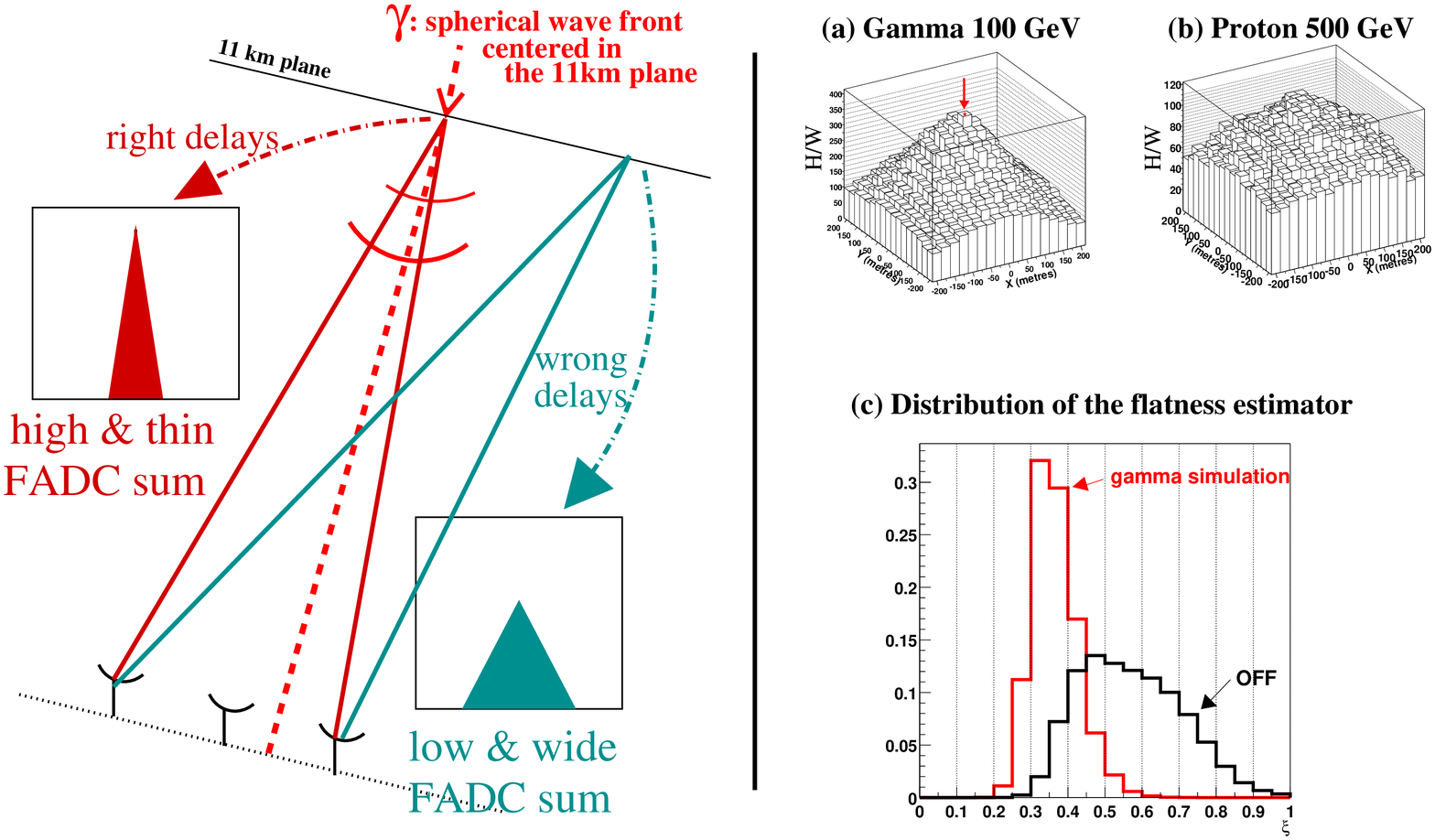}
  \caption{Left: the shape of the FADC sum depends on the position of the centre of the wavefront we assume. Right: views of the height-over-width ratio (H/W) computed for discrete positions in the 11~km plane (a) for a 100 GeV $\gamma$~ray and (b) for a 500 GeV proton. (c) Distribution of the $\xi$ parameter for $\gamma$-rays and OFF data.}
   \label{fig:Timing}
\end{figure}

To detect low energy $\gamma$-rays, we use the sum of the individual digitized signals to increase the signal-to-background ratio. The summation includes a correction for the sphericity of the wavefront, assumed to be centered in the 11~km plane. Assuming a wrong position for this center (impact parameter) broadens the sum: the height-over-width ratio, $(H/W)$, decreases. We compute $(H/W)$ for different assumed positions. The impact parameter is the position for which the $(H/W)$ ratio is maximum, denoted by $(H/W)_{max}$. This is valid for $\gamma$-rays (figure~\ref{fig:Timing}~(a)) but not for protons for which the wavefront is not spherical (figure~\ref{fig:Timing}~(b)). A measurement of the flatness of these 2D-distributions is given by the following estimator:
$$\xi = \mathrm{average}\,\left(\frac{\displaystyle(H/W)_{200m}}{\displaystyle(H/W)_{max}}\right)_{\mbox{\small over 24 positions}},$$
where $(H/W)_{200m}$ is the average of $(H/W)$ over 24 positions along a ring 200~m from the maximum position. For $\gamma$-rays there is a clear maximum and this estimator takes low values. For hadrons, it is usually larger (figure~\ref{fig:Timing} (c)). We have applied $\xi<0.35$. 

\begin{figure}[htbp]
   \centering\includegraphics[width=7.0cm]{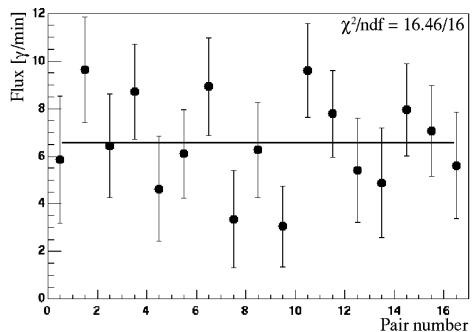}
   \caption{Stability of the Crab detection after hour angle efficiency corrections. Line is $6.5\pm0.5\ \gamma/$min.}
   \label{fig:Crab}
\end{figure}

A cut on this estimator is the biggest contribution to an improvement in Crab sensitivity from 2.2~$\sigma/\sqrt{\mathrm{h}}$ to 5.8~$\sigma/\sqrt{\mathrm{h}}$. The Crab detection is also stable (figure~\ref{fig:Crab}). 

\section{Blazar observations} \label{sec:blazars}

\begin{figure}[htbp]
\begin{center}
   \includegraphics[width=6.0cm]{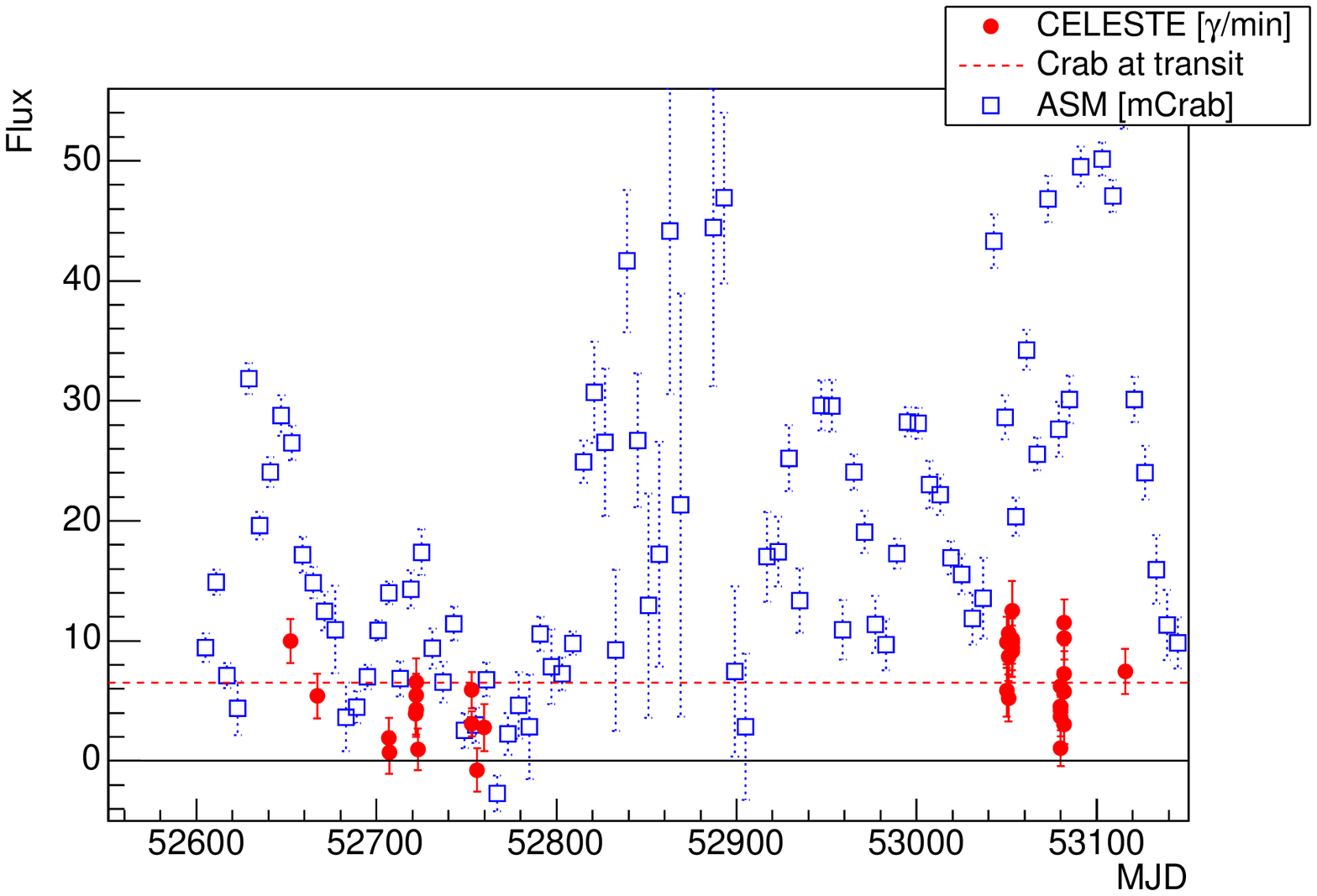}
   \includegraphics[width=6.0cm]{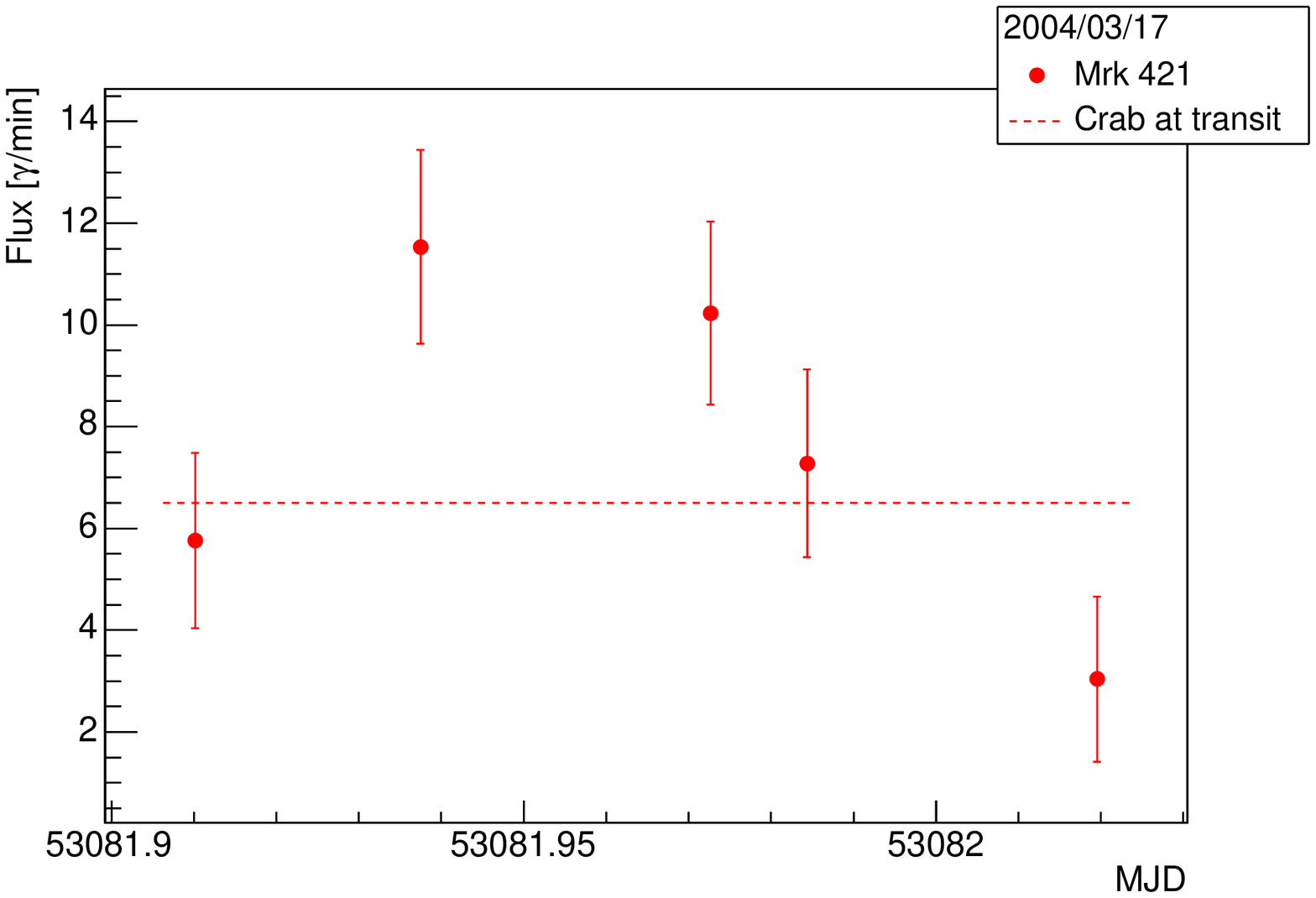}
\end{center}
   \caption{Light curve of Mrk~421. Left: since 2002 seen by CELESTE (red points in $\gamma$/min corrected for hour angle efficiency, 1~point is 20~min data, Crab transit rate shown for reference) and RXTE/ASM (blue squares in mCrab, 6~day bins). Right: light curve of Mrk~421 zoomed on the $17^{\mathrm{th}}$ of March 2004 flare seen by CELESTE.}
   \label{fig:Mrk421}
\end{figure}

This stable analysis and good sensitivity provide a good detection of the Mrk~421 flares. A 19~$\sigma$ detection during 10~h since October 2002 gives a mean of 5.6~$\gamma/$min. The light curve is presented in figure~\ref{fig:Mrk421}. With the new acceptances we will deduce a spectral measurement for Mrk~421 with smaller uncertainties.

The source 1ES~1426+428 is still not detected with our new analysis during the March 2004 observation for 4.4~h data. Since the data have been taken in the same conditions as for the Crab, we can use the sensitivity of 5.8~$\sigma/\sqrt{\mathrm{h}}$ to remark that for a 3~$\sigma$ observation for these 4.4~h data, the flux~$f_\mathrm{1ES\,1426}$ would have to be $f_\mathrm{1ES\,1426} = \frac{3\sigma}{5.8\sigma\sqrt{4.4}}f_\mathrm{Crab} \approx \frac{f_\mathrm{Crab}}{4}$. A better upper limit determination is in preparation.

Finally, the blazar Mrk~501 has been observed during 14.5~h with the pre\-vious experiment at 40 heliostats. The old analysis gave a 2.5~$\sigma$ significance (with 850~photons which would give 1~$\gamma/$min). The new analysis will allow an improved investigation of Mrk~501's behaviour in this unexplored energy range.




\end{document}